\documentstyle[12pt,axodraw,epsf,epsfig]{article}

\newcommand{\beq}{ \begin{eqnarray} }
\newcommand{\eeq}{ \end{eqnarray} }
\newcommand{\beqstar}{ \begin{eqnarray*} }
\newcommand{\eeqstar}{ \end{eqnarray*} }
\newcommand{\gsim}{ \mathop{}_{\textstyle \sim}^{\textstyle >} }
\newcommand{\lsim}{ \mathop{}_{\textstyle \sim}^{\textstyle <} }

\begin{document}
\baselineskip 0.7cm

\begin{titlepage}

\begin{center}

\hfill ICRR-Report-501-203-5\\
\hfill DPNU-03-20\\
\hfill \today

{\large 
$B\rightarrow \phi K_s$ versus Electric Dipole Moment of $^{199}$Hg Atom \\
in Supersymmetric Models with Right-handed Squark Mixing}
\vspace{1cm}

{\bf Junji Hisano}$^{1}$  and 
{\bf Yasuhiro Shimizu}$^{2}$
\vskip 0.15in
{\it
$^1${ICRR, University of Tokyo, Kashiwa 277-8582, Japan }\\
$^2${Department of Physics, Nagoya University, Nagoya 464-8692, Japan}\\
}
\vskip 0.5in

\abstract{
The correlation between the CP asymmetry in $B\rightarrow \phi K_s$
($S_{\phi K_s}$) and the chromoelectric dipole moment (CEDM) of
strange quark ($d_s^C$), which is constrained by the electric dipole
moment (EDM) of $^{199}$Hg, is studied in the supersymmetric (SUSY)  models
with the right-handed squark mixing. It is known that, if the
right-handed bottom and strange squarks have a CP-violating mixing,
such as in the SUSY SU(5) GUT with right-handed neutrinos, the induced
gluon-penguin diagram may give a sizable contribution to $S_{\phi
K_s}$. However, when the left-handed bottom and strange squarks also
have a mixing, the conspiracy of the left and right-handed squarks may
lead to a sizable $d_s^C$, which is enhanced by $m_b/m_s$. While the
estimate for the EDM of $^{199}$Hg, induced by $d^C_s$, might have
large uncertainties due to the hadron and nuclear dynamics, the
current bound implies the gluon penguin contribution by the
right-handed squarks to $S_{\phi K_s}$ should be suppressed and the
deviation of $S_{\phi K_s}$ from the Standard Model may not be so
large.  Also, we discuss the constraint from the EDM of $^{199}$Hg
in the SUSY SU(5) GUT with the right-handed neutrinos.
}

\end{center}
\end{titlepage}
\setcounter{footnote}{0}

The CP violation in $B\rightarrow \phi K_s$ is sensitive to the new
physics since $b\rightarrow s\bar{s}s$ is a radiative process \cite{gw}.
Recently, the Belle experiment in the KEK $B$ factory reported that
the CP asymmetry in $B\rightarrow \phi K_s$ $(S_{\phi K_s})$ is $-0.96\pm
0.50^{+0.09}_{-0.11}$, and $3.5 \sigma$ deviation from the
Standard-Model (SM) prediction $0.731\pm0.056$ is found
\cite{Abe:2003yt}. At present the Babar experiment does not observe such a 
large deviation as $0.45\pm0.43\pm0.07$ \cite{babar}.  Thus, the
combined result is not yet significant, however, the Belle's result
might be a signature of the new physics.

It is known that the supersymmetric (SUSY) models may predict a sizable
deviation of the CP violation in $B\rightarrow \phi K_s$ from the SM
prediction. If the right-handed bottom and strange squarks have a
sizable mixing, the gluon penguin diagram may give a non-negligible
contribution to $b\rightarrow s\bar{s}s$ in a broad parameter space
where the contribution to $b\rightarrow s\gamma$ is a sub-dominant.
The right-handed squark mixing is well-motivated in the SUSY SU(5) GUT
with the right-handed neutrinos, since the tau neutrino Yukawa
coupling may induce the large mixing between the right-handed bottom
and strange squarks \cite{Moroi:2000tk}. Nowadays, $B\rightarrow \phi
K_s$ in the SUSY models is extensively studied \cite{iroiro}\cite{hlmp}.

In this paper the correlation between the CP asymmetry in
$B\rightarrow \phi K_s$ ($S_{\phi K_s}$) and the chromoelectric dipole
moment (CEDM) of strange quark ($d_s^C$) is studied in the SUSY models
with the right-handed squark mixing. In various SUSY models, the
left-handed bottom and strange quark mixing is as large as
$\lambda^2\sim 0.04$ since it is induced via the radiative correction
by the quark Yukawa coupling with the CKM mixing. In this case the
right-handed and left-handed squark mixings between the second and
third generations may lead to the non-vanishing CEDM of the strange
quark. Thus, $S_{\phi K_s}$ and $d_s^C$ may have a strong correlation
in the SUSY models with the right-handed squark mixing. Since $d_s^C$
is constrained by the EDM of $^{199}$Hg, the gluon penguin
contribution from the right-handed squark mixing to $S_{\phi K_s}$
should be suppressed. 

First, we review the constraint on the CEDM of the strange quark.
For the detail, see Ref.~\cite{Falk:1999tm}. The effective Hamiltonian for
the CEDM for quarks is given by
\begin{eqnarray}
 H=  \sum_{q=u,d,s} d_q^C \frac{i}{2}g_s\overline{q}\sigma^{\mu\nu}T^A\gamma_5 q G^A_{\mu\nu}.
\end{eqnarray}
It is known that the CEDMs of the light quarks generate the T-odd
nuclear force, $\bar{N}N\bar{N'}i\gamma_5 N'$, via the T-odd
interaction of $\pi^0$ and $\eta$ to nucleons. For the EDM of
$^{199}$Hg, the EDMs of the constituent nucleons are electrically
screened and the dominant contribution comes from the T-odd nuclear
force
\cite{Flambaum:1984fb}. Now the CEDM of the light quarks are strongly
constrained by measurement of the EDM of $^{199}$Hg atom as
\cite{Romalis:2000mg}
\begin{eqnarray}
e|d_d^C-d_u^C-0.012 d_s^C|&<&7\times 10^{-27}e{\rm cm},
\label{exphg}
\end{eqnarray}
based on the QCD sum rule calculation in Ref.~\cite{Falk:1999tm}. The
coefficient for the CEDM of the strange quark is suppressed compared
with those for CEDM of the down and up quarks, $d_d^C$ and $d_u^C$,
due to the heavier $\eta$ mass and the smaller T-even $\eta$ coupling
than those of pion. From Eq.~(\ref{exphg}),
\begin{eqnarray}
e |d_s^C| &<& 5.8\times 10^{-25}e {\rm cm},
\label{scedm}
\end{eqnarray}
if $d_d^C$ and $d_u^C$ are negligible.  In this paper we use this
constraint on $d_s^C$.  In the theoretical estimate of
Eq.~(\ref{exphg}), the uncertainty from the nuclear dynamics is
dominant, and it might be large.  The neutron EDM may also lead to the
constraint on $d^C_s$ comparable to Eq.~(\ref{scedm}) though it
depends on the calculation \cite{nedm}.  Also, the authors in
Ref.~\cite{Pospelov:2000bw} show from the QCD sum rule that the
Peccei-Quinn symmetry suppresses the contribution of $d_s^C$ to the
neutron EDM.  Thus, we do not use the constraint from the neutron EDM
in this paper.

In the SUSY models, when the left-handed and right-handed squarks have
mixings between the second and third generations, the CEDM of the strange
quark is generated by a diagram in Fig.~1(a), and it is enhanced by
$m_b/m_s$. Using the mass insertion technique, $d_s^C$ is given as
\begin{eqnarray}
  d_s^C &=& c\frac{\alpha_s}{4\pi} \frac{m_{\tilde{g}}}{m^2_{\tilde{q}}}
\left(-\frac{1}{3} N_1(x)-3 N_2(x)\right){\mathrm{Im}}
\left[(\delta_{LL}^{(d)})_{23}
\,(\delta_{LR}^{(d)})_{33}\,(\delta_{RR}^{(d)})_{32}\right],
\label{cedmds}
\end{eqnarray}
where $m_{\tilde{g}}$ and $m^2_{\tilde{q}}$ are the gluino and
averaged squark masses and $c$ is the QCD correction.  We take $c=
3.3$. The functions $N_i$ are given as
\begin{eqnarray}
N_1(x)&=&\frac{3+44x-36x^2-12x^3+x^4+12x(2+3x)\log x}{(x-1)^6},
\\
N_2(x)&=&-2  \frac{ 10 + 9x -18 x^2-x^3+3(1+6x+3x^2) \log x}{(x-1)^6}.
\end{eqnarray}
The mass insertion parameters $(\delta_{LL}^{(d)})_{23}$,
$(\delta_{RR}^{(d)})_{32}$, and $( \delta_{LR}^{(d)})_{33}$ are given by
\begin{eqnarray}
(\delta_{LL}^{(d)})_{23}=
\frac{\left(m_{\tilde{d}_L}^2\right)_{23}}{m^2_{\tilde{q}}},\,
(\delta_{RR}^{(d)})_{32}=
\frac{\left(m_{\tilde{d}_R}^2\right)_{32}}{m^2_{\tilde{q}}},\,
( \delta_{LR}^{(d)})_{33} 
= \frac{m_b\left(A_b -\mu\tan\beta\right)}{m^2_{\tilde{q}}},
\end{eqnarray}
where $(m_{\tilde{d}_{L(R)}}^2)$ is the left-handed
(right-handed) squark mass matrix.  In the typical SUSY models,
$(\delta_{LL}^{(d)})_{23}$ is $O(\lambda^2)\simeq 0.04$. 
>From this formula, $d_s^C$ is estimated in a limit of $x\rightarrow 1$ as
\begin{eqnarray}
  e d_s^C &=&
e c \frac{\alpha_s}{4\pi} \frac{m_{\tilde{g}}}{m^2_{\tilde{q}}}
\left(-\frac{11}{30}\right){\mathrm{Im}}\left[( \delta_{LL}^{(d)})_{23}
\,(\delta_{LR}^{(d)})_{33}\,(\delta_{RR}^{(d)})_{32}\right]
\label{dsap}
\\
&=& -4.0 \times 10^{-23} \sin\theta\, {e\, {\rm cm}} \,
\left(\frac{m_{\tilde{q}}}{500\mathrm{GeV}}\right)^{-3}
\left(\frac{(\delta_{LL}^{(d)})_{23}}{0.04}\right)
\left(\frac{(\delta_{RR}^{(d)})_{32}}{0.04}\right)
\left(\frac{\mu \tan\beta}{5000\mathrm{GeV}}\right)
\end{eqnarray}
where $\theta={\mathrm{arg}}[( \delta_{LL}^{(d)})_{23}
\,(\delta_{LR}^{(d)})_{33}\,(\delta_{RR}^{(d)})_{32}]$. Here,
we neglect the contribution proportional to $A_b$ since it 
is sub-dominant. From this formula, it is obvious that the 
right-handed squark mixing or the CP violating phase should 
be suppressed. For example, for $m_{\tilde{q}}=500$GeV,
$\mu \tan\beta=5000$GeV, and  ${(\delta_{LL}^{(d)})_{23}}={0.04}$,
\begin{eqnarray}
|\sin\theta (\delta_{RR}^{(d)})_{32}| <5.8\times 10^{-4}.
\end{eqnarray}

In Ref.~\cite{bhs} the neutron EDM is discussed in the SUSY SO(10)
GUT, in which the left-handed and right-handed squark mixings are
induced. They show that the EDM of the down quark is enhanced by
$m_b/m_d$ due to the diagram similar to Fig.~1(a).

Now, let us discuss the correlation between $d_s^C$ and $S_{\phi K_s}$
in the SUSY models with the right-handed squark mixing. As mentioned
above, the right-handed bottom and strange squark mixing may lead to
the sizable deviation of $S_{\phi K_s}$ from the SM prediction by the
gluon penguin diagram.  The box diagrams induced by the right-handed
squark mixing also contribute to $S_{\phi K_s}$, however, they tend to
be sub-dominant and do not derive the large deviation of $S_{\phi
K_s}$ from the SM prediction.  Thus, we neglect the box contribution
in this paper for simplicity. The effective operator inducing the
gluon penguin diagram by the right-handed squark mixing is
\begin{eqnarray}
 H&=& - C_8^{R} \frac{g_s}{8\pi^2}
    m_b\overline{s_R}\sigma^{\mu\nu}T^A b_L G^A_{\mu\nu}.
\end{eqnarray}
When the right-handed squarks have the mixing, the dominant
contribution to $C_8^{R}$ is supplied by a diagram with the double
mass insertion of $(\delta_{RR}^{(d)})_{32}$ and
$(\delta_{RL}^{(d)})_{33}$ (Fig.~1(b)). Especially, it is significant
when $\mu\tan\beta$ is large. The contribution of Fig.1~(b) to $C_8^{R}$ 
is  given as
\begin{eqnarray}
 C_8^{R}&=&\frac{\pi \alpha_s}{m^2_{\tilde{q}}}
\frac{m_{\tilde{g}}}{m_b}
(\delta_{LR}^{(d)})_{33}(\delta_{RR}^{(d)})_{32}
(-\frac{1}{3} M_1(x)-3 M_2(x))
\end{eqnarray}
up to the QCD correction. Here,
\begin{eqnarray}
 M_1(x)&=&\frac{1+9x-9x^2-x^3+ (6x+6x^2)\log x}{(x-1)^5},
\\
 M_2(x)&=&-2\frac{3-3x^2+(1+4x+x^2)\log x}{(x-1)^5}.
\end{eqnarray}
In a limit of $x\rightarrow 1$, $C_8^{R}$ is reduced to
\begin{eqnarray}
 C_8^{R}&=&\frac{7\pi \alpha_s}{30{m_b} m_{\tilde{q}}}
(\delta_{LR}^{(d)})_{33}
(\delta_{RR}^{(d)})_{32}.
\label{c8ap}
\end{eqnarray}
Comparing Eq.~(\ref{dsap}) and Eq.~(\ref{c8ap}), we find 
a strong correlation between $d_s^C$ and $C_8^R$ as
\begin{eqnarray}
d_s^C &=& -\frac{m_b}{4\pi^2} \frac{11}{7}
{\mathrm{Im}}\left[( \delta_{LL}^{(d)})_{23} C_8^{R}\right]
\label{massin}
\end{eqnarray}
up to the QCD correction. The coefficient $11/7$ in Eq.~(\ref{massin})
changes from 3 to 1 for $0<x<\infty$.

In Fig.~2, we show the correlation between $d_s^C$ and $S_{\phi K_s}$
assuming a relation $d_s^C = -{m_b}/(4\pi^2) {\mathrm{Im}}[(
\delta_{LL}^{(d)})_{23}C_8^{R}]$ up to the QCD correction. Here, we
take $(
\delta_{LL}^{(d)})_{23} =-0.04$, ${\mathrm{arg}}[C_8^{R}]=\pi/2$ and 
$|C_8^R|$ corresponding to $10^{-5}<|(\delta_{RR}^{(d)})_{32}|<0.5$. The
matrix element of chromomagnetic moment in $B\rightarrow \phi K_s$ is
\begin{eqnarray}
\langle \phi K_S|\frac{g_s}{8\pi^2}m_b(\bar{s}_i \sigma^{\mu\nu}T^a_{ij}P_Rb_j)
G^a_{\mu\nu}| \overline{B}_d\rangle &=&
\kappa \frac{4\alpha_s}{9\pi}(\epsilon_\phi p_B)f_\phi m_\phi^2 F_+(m_\phi^2),
\label{kappa}
\end{eqnarray}
and $\kappa=-1.1$ in the heavy-quark effective theory \cite{hlmp}.
Since $\kappa$ may suffer from the large hadron uncertainty, we take
$\kappa=-1$ and $-2$. From this figure, it is found that the deviation
of $S_{\phi K_s}$ from the SM prediction due to the gluon penguin
contribution should be tiny when  the constraint on $d_s^C$ in
Eq.~(\ref{scedm}) is applied. If we loosen the CEDM constraint by a factor of
100 or more, the large deviation of $S_{\phi K_s}$ by 
the gluon penguin diagram with the right-handed squark mixing
is possible.

We ignored the constraint from $b\rightarrow s \gamma$ in Fig.~2.  If
$(\delta_{RR}^{(d)})_{23}\sim O(1)$, the contribution may  not be negligible.
When the gluino penguin diagrams proportional to
$(\delta_{RR}^{(d)})_{23}$ are dominant,
$Br(B \to X_s \gamma)$ is approximately given as 
\begin{eqnarray}
Br(B \to X_s \gamma) &=& 7.0 \times 10^{-6} 
\left(\frac{\mu \tan\beta}{5000 GeV}\right)^2
\left(\frac{m_{\tilde q}}{500 GeV}\right)^{-4}
\left(\frac{|(\delta_{RR}^{(d)})_{23}|}{0.04}\right)^2 
\label{brbsg}
\end{eqnarray}
for large $\tan\beta$\footnote{This equation is derived from the formula
in Ref.~\cite{masiero}.}.  
Here, we ignore the SM and other SUSY contributions for simplicity.
Imposing that the SUSY contribution in Eq.(\ref{brbsg}) 
does not exceed the central value of the experimental
branching ratio, $Br(B \to X_s \gamma)=(3.3\pm 0.4)\times 10^{-4}$
\cite{pdg},
we can put the experimental bound on 
$(\delta_{RR}^{(d)})_{23}$ as
\begin{equation}
\left|(\delta_{RR}^{(d)})_{23}\right|
\lsim
0.27
\left(\frac{\mu \tan\beta}{5000 GeV}\right)^{-1}
\left(\frac{m_{\tilde q}}{500 GeV}\right)^{2},
\label{bsgconstraint}
\end{equation}
and then, $S_{\phi K_s} \gsim (0.2-0.3)$ $(-(0.3-0.4))$ for $\kappa
=-1(-2)$. 
While this estimate of the bound on $(\delta_{RR}^{(d)})_{23}$ is a
little rough, it is obvious that the contraint from the CEDM of the
strange quark is much stronger than it.

Here, we notice that the above analysis of the CEDM of the strange
quark and $S_{\phi K}$ is changed if the strange quark mass,
$m_s$, is radiatively induced by the one-loop SUSY diagrams and the
tree-level contribution is subdominant.  This is because we have to
perform a chiral rotation of the strange quark to make the mass term
canonical when the induced mass has a phase. However, the $Br(B\to
X_s\gamma)$ constraint implies that the radiative contribution to
$m_s$ should be suppressed. When both the left-handed and right-handed
squarks have  mixings, the dominant SUSY contribution to $m_s$ comes
from Fig.~3 and it is given as
\begin{eqnarray}
 \delta m_s \simeq 3\, {\mathrm MeV} 
\left(\frac{(\delta_{LL}^{(d)})_{23}}{0.04}\right)
\left(\frac{(\delta_{RR}^{(d)})_{32}}{0.04}\right)
\left(\frac{\mu m_{\tilde g}}{m_{{\tilde d}}^2}\right)
\left(\frac{\tan\beta}{50}\right)
\left(\frac{m_b}{5\, \mathrm{GeV}}\right)
\left(\frac{f(x)}{1/3}\right),
\end{eqnarray}
where $f(x)=(2+3 x -6x^2+x^3 +6 x\log x)/6/(1-x)^4$. Thus, when the
left-handed squark mixing is determined by the KM matrix as
${(\delta_{LL}^{(d)})_{23}}\simeq {0.04}$, the one-loop correction to
$m_s$ cannot dominate over the tree level due to the $Br(B\to X_s
\gamma)$ constraint in Eq. (\ref{bsgconstraint}). 
There is the logical possibility that there is  some cancellation 
between the contributions to $Br(B\to X_s \gamma)$, 
in which case the contribution to the strange quark mass may 
not be small. Even in such a case, the effects on $S_{\phi K_s}$ 
is still small. The reason is that the quark mass matrix must be 
re-diagonalized, and the rotation of the right-handed strange quark 
can remove the contributions to $S_{\phi K_s}$.

Now we showed the strong constraint on $S_{\phi K_s}$ from $d_s^C$ in
a case that the right-handed bottom and strange squarks have a mixing.
Let us show the loopholes in this argument. First, $d^C_s$ is induced
by both the left-handed and right-handed squark mixings between the
second and third generations. Thus, if the left-handed squark mixing
$(\delta_{LL}^{(d)})_{23}$ is smaller than $\sim 10^{-4}$, the  constraint
from the CEDM of the strange
quark is not significant. Also, if the left-handed strange squark
is heavier than the other squarks, the $d^C_s$ is suppressed while
$S_{\phi K_s}$ is not changed.  Second, we neglect the CEDM
contribution to the EDM of $^{199}$Hg atom from the up and down quarks
since it depends on the detail of the model. Thus, it might be
possible that the combination $d_d^C-d_u^C-0.012 d_s^C$ is
accidentally canceled bellow the experimental bound. Third, the
estimate for the EDM of $^{199}$Hg atom might have large uncertainly
due to the hadron and nuclear dynamics.  Also, if $\kappa$ in
Eq.~(\ref{kappa}) is extremely large, the sizable deviation of
$S_{\phi K_s}$ might is possible. Fourth, the large left-handed squark
mixing may supply the large gluon penguin contribution 
to $S_{\phi K_s}$ even if the
right-handed squark mixing is suppressed. However, it that case,
$b\rightarrow s \gamma$ constraint is strong, and the cancellation
among the SUSY diagrams in $b\rightarrow s \gamma$ is required.

Finally, we discuss about the constraint from the the EDM of $^{199}$Hg
in the SUSY SU(5) GUT with the right handed neutrinos. In the model
the tau neutrino Yukawa coupling induces the right-handed down-type
squark mixing between the second and third generations radiatively as
\begin{eqnarray}
(m_{\tilde{d}_R}^2)_{32}  &\simeq&-\frac{2}{(4\pi)^2} 
{\rm e}^{i(\varphi_{d_2}-\varphi_{d_3})}
U_{33} U^\star_{23} 
\frac{m_{\nu_\tau} M_{\nu_\tau}}{\langle H_2 \rangle^2}
(3m_0^2+A_0^2) 
\log\frac{M_G}{M_{GUT}},
\label{gut}
\end{eqnarray}
and the mixing is enhanced by the large angle of the atmospheric
neutrino.  Here, we assume for simplicity that the right-handed
neutrino mass matrix is diagonal. $m_{\nu_\tau}$ and $M_{\nu_\tau}$
are the left-handed and right-handed tau neutrino masses, $U$ is the MNS
matrix, and ${M_G}$ and $M_{GUT}$ are the reduced Planck mass and GUT
scale.  The CP violating phase ${\rm
e}^{i(\varphi_{d_2}-\varphi_{d_3})}$ is inherent in the SUSY SU(5)
GUT \cite{Moroi:2000tk}. Since the phase of $U_{33} U^\star_{23}$ is
suppressed due to the small $U_{13}$, the phase of
$(m_{\tilde{d}_R}^2)_{32}$ comes from the GUT inherent one
dominantly. $(m_{\tilde{d}_R}^2)_{32}$, and then $d_s^C$, are
proportional to $M_{\nu_\tau}$. From Eq.~(\ref{gut}),
\begin{eqnarray}
(\delta_{RR}^{(d)})_{32}
&\simeq&-1\times 10^{-3}\times{\rm e}^{i(\varphi_{d_2}-\varphi_{d_3})}
\nonumber\\
&&
\times \left(\frac{m_{\nu_\tau}}{5 \times 10^{-2}{\rm eV}}\right)
\left(\frac{M_{\nu_\tau}}{10^{13}{\rm GeV}}\right)
\left(\frac{U_{33} U^\star_{23}}{1/2}\right)
\left(\frac{3m_0^2+A_0^2}{3 m^2_{\tilde{q}}}\right).
\end{eqnarray}
The CEDM of the strange quark is larger than the experimental bound
when $M_{\nu_\tau}$ is larger than about $10^{12-13}$ GeV and
$(\varphi_{d_2}-\varphi_{d_3})$ is of the order of 1.  This means that
the measurement of the EDM of $^{199}$Hg atom is very sensitive to the
right-handed neutrino sector in the SUSY SU(5) GUT.
The current experimental bound on the EDM of
$^{199}$Hg atom is determined by the statistics, and the further
improvement is expected
\cite{Romalis:2000mg}.

\underline{Acknowledgment}\\
We thank Prof. Murayama for useful discussion about a case that
the strange quark mass is induced radiatively.
This work is supported in part by the Grant-in-Aid for Science
Research, Ministry of Education, Science and Culture, Japan
(No.15540255, No.13135207 and No.14046225 for JH).

\newpage
%
%
\newcommand{\Journal}[4]{{\sl #1} {\bf #2} {(#3)} {#4}}
\newcommand{\APJ}{Ap. J.}
\newcommand{\CJP}{Can. J. Phys.}
\newcommand{\MPL}{Mod. Phys. Lett.}
\newcommand{\NC}{Nuovo Cimento}
\newcommand{\NP}{Nucl. Phys.}
\newcommand{\PL}{Phys. Lett.}
\newcommand{\PR}{Phys. Rev.}
\newcommand{\PRep}{Phys. Rep.}
\newcommand{\PRL}{Phys. Rev. Lett.}
\newcommand{\PTP}{Prog. Theor. Phys.}
\newcommand{\SJNP}{Sov. J. Nucl. Phys.}
\newcommand{\ZP}{Z. Phys.}

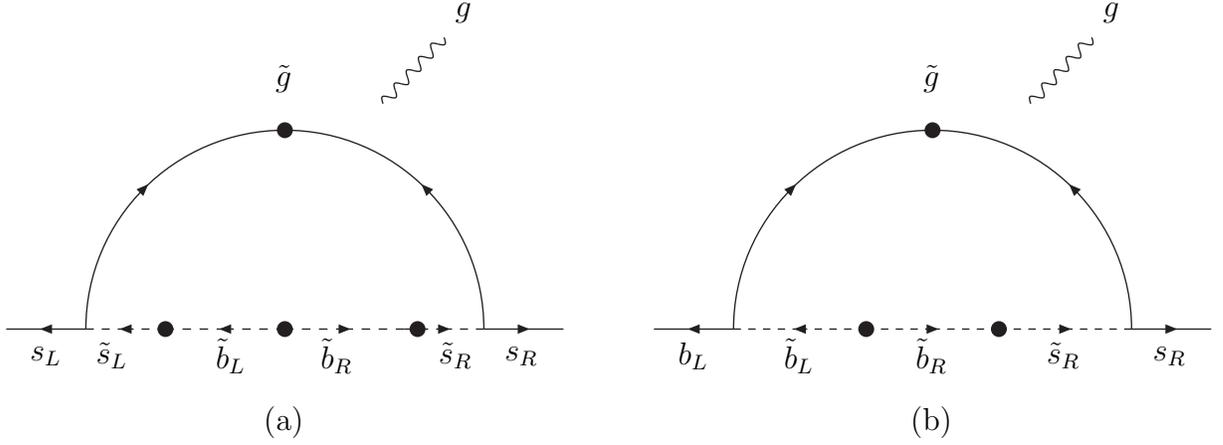
\begin{figure}[p]
\begin{center} 
\begin{picture}(455,140)(30,-20)
\ArrowArcn(135,25)(75,180,90)
\ArrowArc(135,25)(75,0,90)
\Vertex(135,100){3}
\Text(135,120)[]{$\tilde{g}$}

\ArrowLine(60,25)(30,25)
\DashArrowLine(90,25)(60,25){3}             \Vertex(90,25){3}
\DashArrowLine(135,25)(90,25){3}             \Vertex(135,25){3}
\DashArrowLine(135,25)(180,25){3}       \Vertex(185,25){3}  
\DashArrowLine(180,25)(210,25){3}        
\ArrowLine(210,25)(240,25)

\Text(45,15)[]{$s_L$}
\Text(70,15)[]{$\tilde{s}_L$}
\Text(115,15)[]{$\tilde{b}_L$}
\Text(155,15)[]{$\tilde{b}_R$}
\Text(200,15)[]{$\tilde{s}_R$}
\Text(225,15)[]{$s_R$}

\Photon(172,110)(195,135){2}{5}
\Text(203,145)[]{$g$}

\Text(135,-10)[]{(a)}

\SetOffset(245,0)
\ArrowArcn(135,25)(75,180,90)
\ArrowArc(135,25)(75,0,90)
\Vertex(135,100){3}
\Text(135,120)[]{$\tilde{g}$}

\ArrowLine(60,25)(30,25)
\DashArrowLine(110,25)(60,25){3}             \Vertex(110,25){3}
\DashArrowLine(110,25)(160,25){3}            \Vertex(160,25){3}        
\DashArrowLine(160,25)(210,25){3}           
\ArrowLine(210,25)(240,25)

\Text(45,15)[]{$b_L$}
\Text(85,15)[]{$\tilde{b}_L$}
\Text(135,15)[]{$\tilde{b}_R$}
\Text(185,15)[]{$\tilde{s}_R$}
\Text(225,15)[]{$s_R$}

\Photon(172,110)(195,135){2}{5}
\Text(203,145)[]{$g$}

\Text(135,-10)[]{(b)}

\end{picture} 

\caption{a) The dominant diagram contributing to the CEDM of the strange
quark when both the left-handed and right-handed squarks have mixings.
b) The dominant SUSY diagram contributing to the CP asymmetry in
$B\rightarrow \phi K_s$ when the right-handed squarks have a mixing. }

\end{center}
\end{figure}

\begin{figure}
\centerline{
\epsfxsize = 0.5\textwidth \epsffile{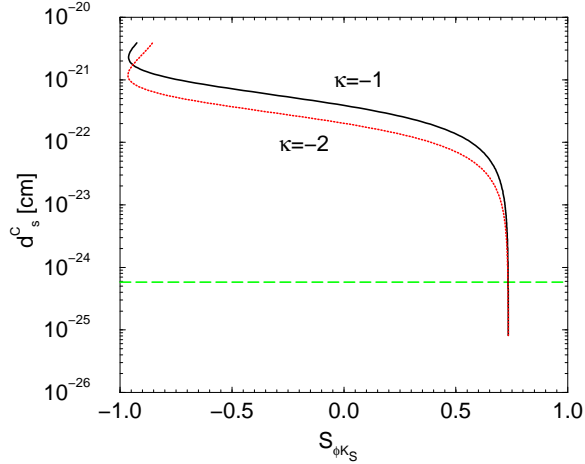} 
}
\vspace*{-5mm}
\caption{
The correlation between $d_s^C$ and $S_{\phi K_s}$ assuming
$d_s^C = -{m_b}/(4\pi^2)
{\mathrm{Im}}[ ( \delta_{LL}^{(d)})_{23}C_8^{R}]$. Here, $( \delta_{LL}^{(d)})_{23}
=-0.04$ and ${\mathrm{arg}}[C_8^{R}]=\pi/2$. $\kappa$ comes from 
the matrix element of chromomagnetic
moment  in $B\rightarrow \phi K_s$.
The dashed line is the upperbound on $d_s^C$ from the EDM of 
$^{199}$Hg atom. 
}
\end{figure}

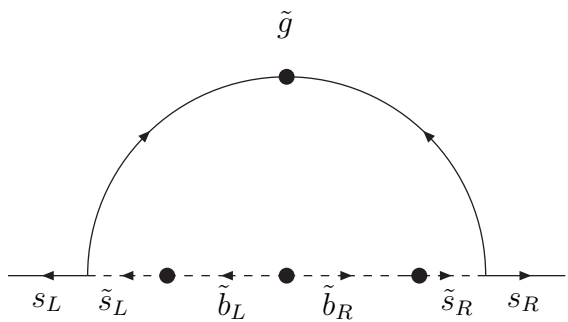
\begin{figure}[p]
\begin{center} 
\begin{picture}(220,140)(30,-20)
\ArrowArcn(135,25)(75,180,90)
\ArrowArc(135,25)(75,0,90)
\Vertex(135,100){3}
\Text(135,120)[]{$\tilde{g}$}

\ArrowLine(60,25)(30,25)
\DashArrowLine(90,25)(60,25){3}             \Vertex(90,25){3}
\DashArrowLine(135,25)(90,25){3}             \Vertex(135,25){3}
\DashArrowLine(135,25)(180,25){3}       \Vertex(185,25){3}  
\DashArrowLine(180,25)(210,25){3}        
\ArrowLine(210,25)(240,25)

\Text(45,15)[]{$s_L$}
\Text(70,15)[]{$\tilde{s}_L$}
\Text(115,15)[]{$\tilde{b}_L$}
\Text(155,15)[]{$\tilde{b}_R$}
\Text(200,15)[]{$\tilde{s}_R$}
\Text(225,15)[]{$s_R$}

\end{picture} 

\caption{The dominant SUSY contribution to  the strange
quark mass when both the left-handed and right-handed squarks have
mixings. }

\end{center}
\end{figure}

\end{document}